\documentclass[twocolumn]{jpsj2} 
\def\runauthor{N. Furukawa and Y. Motome}

\title{Colossal magnetoresistance and quenched disorder in manganese oxides}

\author{Nobuo \textsc{Furukawa}$^{1}$\thanks{E-mail address: furukawa@phys.aoyama.ac.jp} 
and 
Yukitoshi \textsc{Motome}$^{2}$}

\inst{$^{1}$Department of Physics, Aoyama Gakuin University, 
Kanagawa 299-8558, Japan \\
$^{2}$RIKEN (The Institute of Physical and Chemical Research), 
Saitama 351-0198, Japan}

\abst{
We give an overview on several recent topics of colossal magnetoresistive manganites 
in both experiments and theories, focusing on the effect of quenched disorder.
The disorder is intrinsically involved since the compounds are solid solutions, 
and its importance has been pointed out in several experiments of 
transport and magnetic properties. 
Recent progress in the experimental control of the strength of disorder 
is also reviewed. 
Theoretically, the effect of the disorder has been explored
within the framework of the double-exchange mechanism. 
Several efforts to understand the phase diagram and 
the electronic properties are reviewed. 
We also briefly discuss a recent topic on 
the effect of disorder on competing phases and 
the origin of colossal magnetoresistance.
}

\kword{colossal magnetoresistive manganites, quenched disorder, 
double-exchange model, competing orders
}

\begin{document}
\def\TC{T_{\rm C}}
\maketitle

\section{Introduction} 

Colossal magnetoresistance (CMR) in 
perovskite manganese oxides has been one of the central issues 
in condensed matter physics. 
\cite{Ramirez1997,Tokura1999,Coey1999,Salamon2001,Dagotto2001}
CMR is a gigantic response of the electrical resistivity 
to the external magnetic field; 
the resistivity sharply decreases even on the order of 
$10^4$-$10^6$ by applying the magnetic field of only a few Tesla. 
Besides fundamental understanding of the mechanism of CMR, 
possibilities of technological applications have motivated this field. 

CMR manganites exhibit a variety of physical properties 
such as ferromagnetism, antiferromagnetism, metal-insulator transition, 
charge ordering and orbital ordering. 
The most universal phenomenon is 
the transition from the high-temperature paramagnetic phase 
to the low-temperature ferromagnetic metallic phase. 
\cite{Wollan1955,Searle1970}
This transition is basically understood 
by the double-exchange (DE) mechanism. 
\cite{Zener1951,Anderson1955,deGennes1960,Furukawa1999a}
There, $S=3/2$ localized spins in $t_{2g}$ orbitals and 
itinerant electrons in $e_g$ orbitals are strongly coupled 
by the ferromagnetic Hund's-rule coupling, which 
gives rise to the intimate interplay 
between the electron conductivity and the magnetism. 
However, it is obvious that the DE mechanism is insufficient 
to understand the rich phenomena in the CMR manganites comprehensively. 
Therefore, the key issue is how to explain the various properties and
what kind of extension is necessary to the simple DE picture. 

In these CMR manganites, whose chemical formula is generally given by $A$MnO$_3$, 
electronic properties strongly depend on $A$-site cations. 
That is, one may control the physical properties 
by substituting the $A$-site cations.
There are two types of the substitution: 
One is the substitution with different ionic valences, 
which is usually called the doping control. 
An example is a change of Sr concentration $x$ in La$_{1-x}$Sr$_x$MnO$_3$, 
where valence difference between La$^{3+}$ and Sr$^{2+}$ 
effectively introduces mobile holes in Mn sites. 
The other substitution is by different ionic radii at a fixed valence, 
which is often called the bandwidth control. 
An example is Sr substitution to Ca sites in Pr$_{1-x}$Ca$_x$MnO$_3$ with fixing $x$. 
Here the nominal valence of Mn cations is unchanged, 
however, the difference of ionic radius between Sr$^{2+}$ and Ca$^{2+}$ 
modifies local lattice structure, i.e., 
the angle and length of Mn-O-Mn bonds surrounding the cations substituted. 
This modifies the hopping integrals between Mn sites, 
which leads to a change of the effective electron bandwidth. 
Note that the former substitution includes both effects of 
the valence control and the ionic-radius control. 
Fine tuning of these two types of the substitutions enables 
to control drastic changes of physical properties including CMR. 

An important issue in these valence control and ionic-radius control is that 
these substitutions inevitably introduce quenched disorder as well 
since the systems are solid solution compounds. 
Random distribution of different $A$-site cations leads to 
structural and electronic disorder. 
In the long survey of the manganites, 
the disorder effect has not attracted much attention, and 
has been considered as a secondary effect to the doping or bandwidth controls. 
However, this hidden effect has been reconsidered 
in recent experimental and theoretical efforts, and 
revealed to play a crucial role in the physics of manganites. 

In this contribution, we give an overview of the investigation on 
the disorder effect in CMR manganites. 
This is not a comprehensive review 
but may be a (biased) summary of recent trends in this field. 
This paper is organized as follows. 
In the next section 2, we will give short survey of 
experimental indications for the importance of the disorder. 
In Sec.~3, several theoretical efforts are introduced 
to explain the disorder effects in manganites. 
Section~4 is devoted to discussion on a possible mechanism of CMR 
with emphasis on the role of quenched disorder.

\section{Experimental}

In this section, we show several experimental studies 
which indicate the importance of the quenched disorder in CMR manganites.

\subsection{Residual resistivity}

\begin{figure}[t]
\begin{center}
\includegraphics[width=6.5cm]{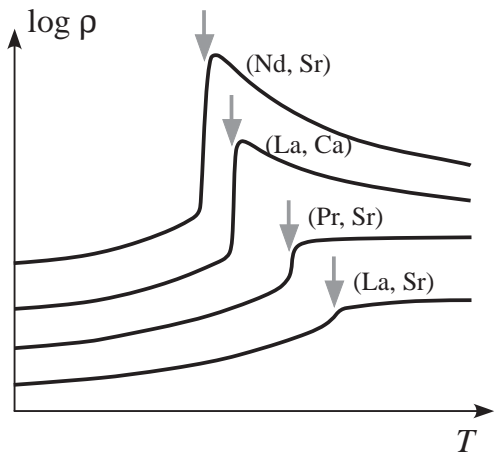}
\end{center}
\caption{
Schematic picture of the temperature dependences of the resistivity 
of $A_{0.7} A'_{0.3}$MnO$_3$. The combinations of ($A, A'$) are shown.
Gray arrows indicate the Curie temperature $T_{\rm C}$.}
\label{f1}
\end{figure}

Disorder effect is apparently found in 
the residual resistivity, i.e., the value of the electrical resistivity 
as the temperature approaches zero. 
Coey {\it et al.} reported the transport properties of $A_{0.7} A'_{0.3}$MnO$_3$ 
for various combinations of $A^{3+}$ and $A'^{2+}$ cations, 
and found that the residual resistivity largely depends on 
the combination of ($A, A'$). 
\cite{Coey1995}
Temperature dependences of the resistivity are schematically shown in Fig.~\ref{f1}. 
These substitutions of the ($A, A'$) cations 
correspond to the ionic-radius control and 
does not change the doping concentration. 
At the same time, the effective mass, 
which is estimated by the specific heat measurement, 
does not show any significant enhancement. 
\cite{Coey1995}
Hence, the change of the residual resistivity demonstrates that 
different ionic radii of the $A$-site cations indeed introduce 
the quenched disorder in the system, 
which scatters itinerant electrons at low temperatures and 
changes the lifetime of quasi-particles. 

The increase of the residual resistivity appears to correlate with
the decrease of the Curie temperature $T_{\rm C}$, 
\cite{Coey1995,Saitoh1999}
as schematically indicated in Fig.~\ref{f1}.
This correlation has been systematically explored, and 
will be discussed in the next section. 

At the same time, as the residual resistivity increases, 
the system becomes more insulating above $T_{\rm C}$, and 
exhibits larger sudden drop of the resistivity at $T_{\rm C}$ 
as shown in Fig.~\ref{f1}. 
The larger drop leads to more sensitive response 
to the external magnetic field, i.e., the larger CMR effect. 
Thus, the insulating behavior above $T_{\rm C}$ is crucial 
to consider the mechanism of CMR.
This issue will be discussed in Sec.~4.

\subsection{Curie temperature}

In the simple DE scenario, the Curie temperature $T_{\rm C}$ 
scales to the effective bandwidth. 
Radaelli {\it et al.} pointed out that this scaling no longer hold 
in the series of the ionic-radius control in $A_{0.7} A'_{0.3}$MnO$_3$. 
\cite{Radaelli1997}
They found that the change of $T_{\rm C}$ is much larger than 
that of the bandwidth. 
For instance, from ($A, A'$) = (La, Sr) to (La, Ca), 
$T_{\rm C}$ decreases by about $30$\% 
while the bandwidth estimated from the structural data 
decreases by less than $2$\%. 
This implies a hidden mechanism to suppress the kinetics of electrons 
in this substitution of the ionic-radius control. 

More direct evidence was revealed by Rodriguez-Martinez and Attfield. 
\cite{Rodriguez-Martinez1996}
They measured $T_{\rm C}$ for $A_{0.7} A'_{0.3}$MnO$_3$ 
with the same average ionic radius of the $A$ sites, 
i.e., the same effective bandwidth. 
They found that even for the same bandwidth, 
$T_{\rm C}$ varies according to the combinations of ($A, A'$) and 
well scales to the variance of the $A$-site ionic radius: 
$T_{\rm C}$ decreases for larger variance. 
This scaling strongly indicates that the disorder due to 
the random distribution of $A$-site cations with different ionic radii 
suppresses the motion of itinerant electrons.

\subsection{Spin excitation anomalies}

Another indication, which is rather indirect, is found 
in the spin excitation spectrum. 
Hwang {\it et al.} reported that the magnon dispersion 
shows softening and broadening near the Brillouin zone boundaries. 
\cite{Hwang1998}
Dai {\it et al.} confirmed that these anomalous behaviors appear universally 
in compounds showing rather low $T_{\rm C}$. 
\cite{Dai2000}
On the contrary, in compounds showing higher $T_{\rm C}$ 
such as La$_{0.7}$Pb$_{0.3}$MnO$_3$, 
the magnon dispersion does not show the anomalies and 
obeys a cosine-like form, 
\cite{Perring1996}
which is well explained by the simple DE model. 
\cite{Furukawa1996}
These results suggest a hidden mechanism beyond DE 
in the low-$T_{\rm C}$ compounds. 


Recently, it has been pointed out that 
these anomalies can be explained simply by introducing 
quenched disorder to the DE model. 
\cite{Motome2002a,Motome2003a,MotomePREPRINT}
This scenario will be discussed in Sec.~3.5.

\subsection{$A$-site order/disorder materials}

A big progress for understanding the disorder effect has been made recently.
Millange {\it et al.} demonstrated that it is possible to synthesize 
mixed valence compounds $A_{0.5} A'_{0.5}$MnO$_3$ 
in which the $A^{3+}$ and $A'^{2+}$ cations order periodically in lattice structure. 
\cite{Millange1998}
These special compounds, which are called the $A$-site ordered manganites, 
exhibit distinct properties from the materials with random configuration 
of $A$-site cations. 
We can compare the `clean' systems ($A$-site ordered) and 
the `dirty' systems ($A$-site disordered) 
both of which have the same chemical formula.

\begin{figure}[t]
\begin{center}
\includegraphics[width=7cm]{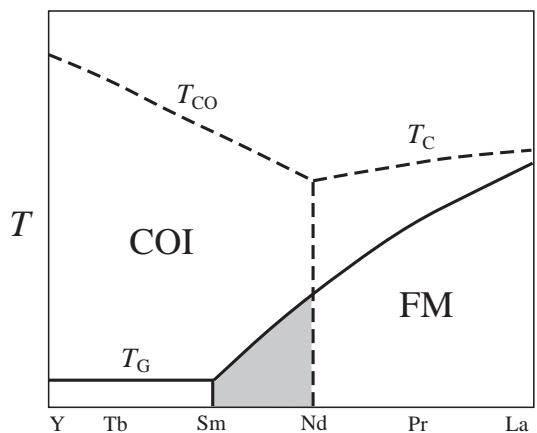}
\end{center}
\caption{
Schematic phase diagram for the $A$-site order/disorder manganites 
$A_{0.5} A'_{0.5}$MnO$_3$ with $A'$=Ba.
$A$-site cations are shown in the horizontal axis. 
Dashed (Solid) curves are the phase boundaries for 
the ordered (disordered) materials: 
$T_{\rm C}$, $T_{\rm CO}$ and $T_{rm G}$ represents 
the transition temperatures for 
the ferromagnetic metallic phase (FM), 
the charge and orbital ordered phase (COI) and 
the spin glass state, respectively. 
The gray region shows the regime where the disorder-induced 
insulator-to-metal transition takes place. 
}
\label{f3}
\end{figure}

One of the most appealing aspects of these $A$-site ordered materials is 
multicritical behavior between the ferromagnetic metallic phase and 
the charge and orbital ordered insulating phase, 
as pointed out by Nakajima {\it et al.} and Akahoshi {\it et al.}
\cite{Nakajima2002,Akahoshi2003,Nakajima2004}
When the disorder is introduced, 
the charge/orbital ordered phase becomes unstable and 
replaced by some glassy phase, and therefore, 
the multicritical point is strongly suppressed 
down to much lower temperature. 
These drastic and asymmetric changes of the phase diagram 
are depicted in Fig.~\ref{f3} schematically. 
More interestingly, in the multicritical region, 
the system shows a disorder-induced insulator-to-metal transition. 
In the disordered compounds, 
charge-order (CO) fluctuations are surprisingly enhanced above $T_{\rm C}$, 
\cite{Tomioka2003}
and a typical CMR behavior is observed. 
\cite{Akahoshi2003}
These results indicate that the competing orders and the quenched disorder 
play an important role in the CMR phenomena.

\section{Theoretical}

Here we pick up several theoretical efforts to understand 
the disorder effect in CMR manganites.

\subsection{Green's function technique and transfer matrix method}

In an early stage, the effect of the quenched disorder 
was examined by using some Green's function technique. 
Allub and Alascio studied by using the renormalized perturbation expansion, 
and discussed the localization of the electrons. 
\cite{Allub1996}
Mazzaferro {\it et al.} applied the equation-of-motion method and 
calculated the energy difference between ferromagnetic state and 
paramagnetic state. 
\cite{Mazzaferro1985}

The localization-delocalization phenomenon has also been examined 
by the transfer matrix method. 
\cite{Sheng1997a,Sheng1997b,Li1997}
By including the quenched disorder and 
the random hoppings of itinerant electrons 
under the influence of the local magnetic field by localized spins, 
the phase diagram as well as transport properties are explored.

\subsection{First-principle calculation}

Pickett and Singh examined the effect of disorder due to 
the $A$-site substitution by using 
the first-principle local density functional method 
with a virtual-crystal approximation. 
\cite{Pickett1997}
An important result is the distribution of 
screened Coulomb energy at Mn sites. 
They showed that the potential energy is indeed disordered and 
its distribution has the width of the order of $0.1$eV. 
This energy scale is comparable to the transfer integrals of 
itinerant electrons, and hence cannot be neglected.

\subsection{Dynamical mean-field study}

Several studies have been done by using 
the dynamical mean-field theory (DMFT). 
\cite{Zhong1998,Letfulov2001,Auslender2001,Narimanov2002}
The DMFT for the DE model in the absence of the disorder 
has successfully reproduced the CMR phenomena 
in rather `clean' systems. 
\cite{Furukawa1994,Furukawa1995a,Furukawa1995b}
The calculations have been extended to the cases with the disorder. 
The density of states, $T_{\rm C}$, and 
transport properties have been calculated as functions of 
the strength of disorder, the doping concentration, and temperature. 
The disorder effect on the ferromagnetic metal transition is 
qualitatively explained by the series of the DMFT studies. 

An important effect which is missed in the DMFT is 
spatial fluctuation, 
while it fully includes the thermal fluctuations. 
\cite{Georges1996}
In real materials, itinerant electrons move around with feeling 
the effect of disorder which are randomly distributed in space. 
The spatial fluctuation may be relevant for more quantitative argument. 
This is indeed the case as we will show in the next section.

\subsection{Monte Carlo study}

Monte Carlo (MC) simulation has also been performed 
for the DE model with the quenched disorder. 
\cite{Motome2002b,Motome2003b}
In this technique, both thermal and spatial fluctuations are included 
and essentially exact results are obtained numerically 
for finite-size clusters. 
It is necessary to perform a large scale simulation 
in order to take many random averages and 
examine the finite-size effect by using the finite-size scaling analysis. 
The authors have developed a new MC algorithm 
which drastically reduces the computational cost. 
\cite{Motome1999,Furukawa2004}

\begin{figure}[t]
\begin{center}
\includegraphics[width=7.5cm]{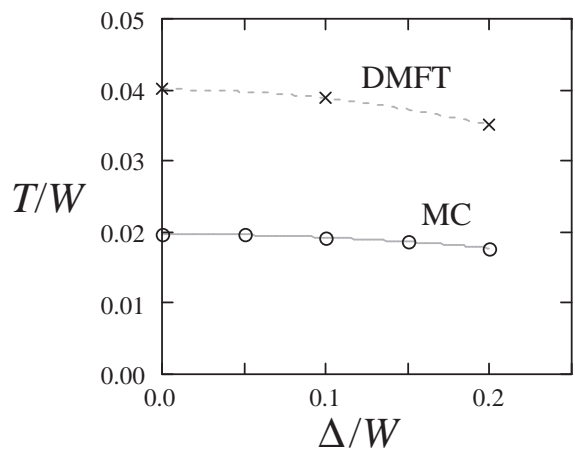}
\end{center}
\caption{
Curie temperature $T_{\rm C}$ as a function of the strength of 
the quenched disorder $\Delta$ for the doping concentration $x=0.3$. 
$W$ is the half bandwidth.
The DMFT results are from Ref.~\ref{Auslender} and 
the MC results are from Ref.~\ref{Motome}. 
Note that in the two studies, the distribution of the disorder is 
commonly the binary one, but in a slight different form.
}
\label{f4}
\end{figure}

The comparison between the DMFT results and the MC results is shown 
in Fig.~\ref{f4}. 
The MC estimates of $T_{\rm C}$ are significantly lower than those by the DMFT, 
which suggests that the spatial fluctuations substantially suppress $T_{\rm C}$.  
\cite{Motome2003b,Motome2000}
The result clearly indicates the importance of the spatial fluctuations. 

It is explicitly shown that 
$T_{\rm C}$ scales to the kinetic energy of electrons, 
which decreases with the square of the strength of disorder 
in the weak disorder regime. 
\cite{Motome2003b}
It is also indicated that 
MC results well reproduce the large reduction of $T_{\rm C}$ 
mentioned in Sec.~2.2 by taking into account 
the strength of disorder estimated 
by the first-principle calculation (Sec.~3.2).

\subsection{Spin excitation anomalies}

The authors pointed out that the spin excitation anomalies described in Sec.~2.3
can be explained by the DE model with the quenched disorder. 
\cite{Motome2002a,Motome2003a,MotomePREPRINT}
The disorder induces broadening, branching 
(softening of the lower branch), and gap-opening 
at around the Fermi wave number $k_{\rm F}$ of the itinerant electrons. 
A typical result is shown in Fig.~\ref{f5}. 
The origin of the anomalies is the Friedel oscillation: 
The itinerant electrons tend to screen the quenched disorder, 
which induces the $2k_{\rm F}$ charge density oscillation. 
This charge density oscillation is equivalent to 
the spin density oscillation because the system is in the half metallic state, 
i.e., the perfectly spin-polarized state.
Thus, the nonmagnetic disorder couples with the magnetic excitation, and 
induces the anomalies at $k_{\rm F}$. 
This novel effect has been recently confirmed in perturbative approaches. 
\cite{Fukui2004,Semba2004}

\begin{figure}[t]
\begin{center}
\end{center}
\caption{
Spin excitation spectrum obtained by the linear spinwave approximation 
for the DE model with the quenched disorder. 
The gray scale shows the intensity of the spectrum.
The dashed curve is the cosine dispersion in the absence of the disorder. 
}
\label{f5}
\end{figure}

Although the disorder scenario appears to give a comprehensive understanding 
of the various properties of CMR manganites, 
several other mechanisms have also been proposed for
the spin excitation anomalies, for instance, 
antiferromagnetic superexchange interactions between the localized spins, 
\cite{Solovyev1999}
orbital degree of freedom in the $e_g$ bands, 
\cite{Khaliullin2000}
the electron-lattice coupling, 
\cite{Furukawa1999b}
and the electron-electron correlation. 
\cite{Kaplan1997,Golosov2000,Shannon2002}
The microscopic origin is still controversial among these scenarios.
Further experiments are desired to settle this controversy.

\subsection{Competing phases and disorder-induced insulator-to-metal transition}

The multicritical behavior in the $A$-site order/disorder manganites (Sec.~2.4) 
has also been studied theoretically 
by explicitly taking account of the quenched disorder. 
\cite{Motome2003c,Aliaga2003,Sen2004}
Here, to describe the competition between the ferromagnetic metal and 
the CO insulator, 
an extended DE model has been studied by incorporating 
the coupling to lattice distortions. 
MC calculation has been applied to include large fluctuation effects 
in this competing disordered system. 

MC results well reproduce the asymmetric response of the phase diagram 
to the quenched disorder. 
The ferromagnetic metallic phase remains robust though $T_{\rm C}$ is reduced. 
On the contrary, the CO phase is surprisingly fragile 
against the disorder. 
As a consequence, the multicritical point shifts to the CO regime and 
strongly suppressed down to lower temperature. 
Numerical results also reproduce the disorder-induced insulator-to-metal transition 
as observed in experiments. 
\cite{Motome2003c,Sen2004}
In this regime, it is clarified that the CO fluctuation, i.e., 
a remnant of the long-range CO which is destroyed by the disorder 
is enhanced toward $T_{\rm C}$ as temperature decreases 
as observed in experiments. 
\cite{Motome2003c}
This leads to insulating behavior of the resistivity, which is 
a key aspect for the typical CMR effect. 

These results reveal that the quenched disorder plays a crucial role 
to induce the CMR in the phase competing regime. 
This gives a clue to enhance the CMR effect and 
open a way of applications in future.

\section{Discussion}

One of the key issues concerning the CMR phenomena
in manganites is to clarify
why  insulating state appears at higher temperatures
while metallic state
exists at lowest temperatures, as mentioned in Sec.~2.1.
Experimentally, charge localizations in manganites are
closely related to CO fluctuations
in such a way that
enhancements of CO fluctuations above $\TC$ suddenly disappear
below $\TC$.\cite{Dai00,Adams00}
In order to investigate the mechanism of the
CMR phenomena, it is crucial to clarify
why energetically unstable charge localized states
arise in the higher temperature range.

Various theories based on polaron mechanisms
claim that localized polarons are formed above $\TC$ to
gain energies through electron-lattice couplings,
while polarons are deformed to gain kinetic energies
in the FM phase below $\TC$.
\cite{Millis1995,Millis1996,Green2001}
These scenarios do not explain why
a reentrant transition from CO fluctuating region
 to FM phase
appears in wide range of the phase diagrams where CMR is observed.

In order to understand this phenomena,
the authors have proposed a scenario
where the reentrant behaviors are driven by the entropy gain
due to the random pinning potentials.\cite{Furukawa05}

Let us consider a system where the CO phase
competes with the FM phase.
By adding quenched disorder, phase of the CO order parameter 
couples with the local pinning potentials.
Since the pinning potentials are randomly distributed,
phase mismatches create the domain structures.
In other words, random pinning potentials introduce
frustrations to the phase of the CO order parameter.
Effects of the frustration is i) a destruction of the CO long range order,
ii) an increase of the energy of the CO state, and
iii) an increase of the entropy of the CO state.

At the same time, phase of the FM order parameter
does not couple with the local pinning potentials.
In the presence of the CO-FM competition,
introduction of the pinning potential destroys
the CO phase and replaces it with the FM phase
at lower temperature range.
On the other hand, at higher temperature range,
the CO fluctuations is stabilized by its large entropy.
This results in an entropy-driven reentrant behavior of 
CO fluctuations relevant to CMR, which 
also explains the phase diagrams of the A-site
ordered/disordered manganites.

Let us now discuss from the viewpoint of
designing  a new material which exhibits CMR.
We consider that it is necessary to introduce an insulating phase
which couples with random pinning potentials
in such a way that frustrations are 
introduced to the phase of the order parameter.
Although the order parameter of the insulating phase
could be anything, it is favorable if
the ordering temperature scale is high enough so that
the CMR phenomena are observed from higher temperatures.

\section*{Acknowledgment}

This work is supported by a Grant-in-Aid and NAREGI 
from the Ministry of Education, Science, Sports, and Culture.

\end{document}